\documentclass[pra, aps, twocolumn, preprint, superscriptaddress, nofootinbib, tightenlines, nobibnotes, showpacs, groupadress]{revtex4-1}
\usepackage{slashed}
\usepackage{amsfonts}
\usepackage{amssymb}
\usepackage{graphicx}
\usepackage{subfigure}
\usepackage[usenames, dvipsnames]{color}
\usepackage{graphics}
\usepackage{bm}
\usepackage{amsmath}
\usepackage{color}
\usepackage{amsfonts}
\usepackage{epsfig}

\everymath{\displaystyle}

\begin{document}
\title{Electron trapping in freely expanding ultracold neutral plasmas}

\author{R. Ayllon}
\affiliation{Instituto de Plasmas e Fus\~ao Nuclear, Lisboa, Portugal}
\affiliation{Instituto Superior T\'ecnico, Lisboa, Portugal}

\author{H. Ter\c{c}as}
\email{hugo.tercas@tecnico.ulisboa.pt}
\affiliation{Instituto de Plasmas e Fus\~ao Nuclear, Lisboa, Portugal}
\affiliation{Instituto Superior T\'ecnico, Lisboa, Portugal}

\author{J. T. Mendon\c{c}a}
\email{titomend@ist.utl.pt}
\affiliation{Instituto de Plasmas e Fus\~ao Nuclear, Lisboa, Portugal}
\affiliation{Instituto Superior T\'ecnico, Lisboa, Portugal}

\begin{abstract}

We report on the self-induced electron trapping occurring in a ultracold neutral plasma that is set to expand freely. At the early stages of the plasma, the ions are not thermalized follow a Gaussian spatial profile, providing the trapping to the coldest electrons. In the present work, we provide a theoretical model describing the
electrostatic potential and perform molecular dynamics simulations to validate our findings. We show that in the strong confinement regime, the plasma potential is of a Thomas-Fermi type, similar to the case of heavy atomic species. The numerically simulated spatial profiles of the particles corroborate this claim. We also extract the electron temperature and coupling parameter from the simulation, so the duration of the transient Thomas-Fermi is obtained. 

\end{abstract}

\maketitle

\section{Introduction}
Owing to the advances performed in laser cooling and trapping of atoms, ultracold neutral plasmas (UNPs) have received an interesting hype in the last decade. The main reason stems in the fact that these systems can be used to investigate strongly coupled matter in a very flexible manner. A striking feature of UNPs are the appreciable high values of the coupling parameter, the ratio between the interaction (Coulomb) and the kinetic (thermal) energies. In addition, UNPs provide a unique platform to study plasma behavior occuring in astrophysical enviroments \cite{Science.316.705}, being experimentally produced via photoionization of a laser-cooled atomic cloud in a magneto-optical trap \cite{PhysRevLett.83.4776, PhysRevLett.85.318, PhysRevLett.86.3759}. In typical experimental conditions, the electrons are produced in the temperature range of $1 - 100$ K \cite{PhysRep.449.77}, while the ions can have a temperatures well below $1$ K \cite{PhysRevA.78.013408, EurPhysLett.89.53001}. The production of UNPs from cold molecular gases has also been reported \cite{PhysRevLett.101.205005, JPhysBAtMolOptPhys.44.184015, JPhysBAtMolOptPhys.47.245701}.\par

The evolution of a freely expanding UNP comprises three different stages, namely electron equilibration \cite{PhysRevLett.99.145001}, ion equilibration \cite{PhysRevLett.92.143001}, and plasma expansion \cite{PhysPlasmas.22.043514}. During the latter, free electrons are able to escape the cloud, producing a electron depletion and thus a net positive charge in the core. This generates a weak electric field that traps the remaining electrons. It has been observed, however, that such local charge imbalance does not affect the self-similar expansion (Coulomb explosion) of the plasma \cite{PhysPlasmas.23.092102}. The ion radial distribution can be obtained by fluorescence techniques \cite{PhysRevLett.95.235001}. However, the electron distribution is not acessible with optical tools, what hinders the access to the dynamical evolution of the electrons. It is, therefore, desirable to investigate eventual signatures of the electrons in the dynamics of the ions. In the present paper, we discus a novel electrostatic regime that can be established at the inner core of expanding UNPs, as a consequence of the electron trapping. We show that a quasi-static equilibrium, resulting from the combination of both trapped and escaping electrons, can be described by the Thomas-Fermi (TF) model, usually employed to described the electronic distribution in heavy atomic species \cite{LandauLifshitz}. We develop a theoretical model that cast the effects of electron trapping and perform a molecular dynamics (MD) simulations to understand the validity regime of the proposed TF model.\par

The paper is organized as follows. In Section II, we present the basic equations leading to the TF model. In Section III, we describe the methods for the MD simulations. A detailed comparison between the theoretical model and the MD simulations are presented in Sec. IV. Finally, in Sec. V, the implications of our findings are discussed and some conclusions are stated. 

\section{The Thomas-Fermi model}

We start with Poisson's equation describing the potential of the UNP cloud
\begin{equation} 
\label{uno_tito}
\nabla^2 \Phi = \frac{e}{n_0}\left( n_e - n_i \right),
\end{equation}
where $n_e$ and $n_i$ represent the electron and ion number density, respectively. Right after the production of the plasma, and assuming the plasma to be produced from the complete photoionization of a cold atomic gas confined in a magneto-optical trap \cite{PhysRep.449.77}, the ions follow a Gaussian profile,
\begin{equation} 
\label{tres_tito}
n_i = n_0 \exp \left( -\frac{r^2}{2\sigma_{0}^{2}} \right),
\end{equation}
with $\sigma_0$ being the size of the plasma cloud. These ions create an electrostatic potential $\Phi > 0$, in such way that the energy of the electrons is given by
\begin{equation} 
\label{cuatro_tito}
E_e(\mathbf{r}, v) = \frac{1}{2}m_ev^2 - e \Phi(\mathbf{r}).
\end{equation}
Due to the electron trap, the energy in \eqref{cuatro_tito} can be negative. As such, trapping occurs for electron velocities satisfying the condition $v < v_{t}$, where the trapping velocity is given by
\begin{equation} 
\label{cinco_tito}
v_t = \sqrt{\frac{2e}{m_e} \left|\Phi\right|}.
\end{equation}
By putting Eqs. (\ref{uno_tito}) and (\ref{tres_tito}) together, and defining $\phi = e \left|\Phi\right| / T_e$, we can write Eq. (\ref{uno_tito}) as
\begin{equation} 
\label{seis_tito}
\nabla^2 \phi = \frac{1}{\lambda_{D}^{2}} \left( \frac{n_e}{n_0} - \exp \left( -\frac{r^2}{2\sigma_{0}^{2}} \right)\right),
\end{equation}
where $\lambda_D = v_{\rm th}/\omega_{pe}$ is the Debye length, $\omega_{pe} = \left(e^2 n_0 /\varepsilon_0 m_e \right)^{1/2}$ is the electron plasma frequency and $v_{th} = \sqrt{T_e /m_e}$ is the electron thermal velocity. While the untrapped electrons - i.e. those with velocity $v_e > v_t$ - follow the Boltzmann distribution associated with (\ref{cuatro_tito}), the trapped ones adiabatically follow the ions, thus being uniformly distributed. We here assume the trapped electrons to be confined within the radius $R \sim \sigma_0$. This assumption is valid for the early stages of the plasma expansion, as we will later confirm with the MD simulations. Therefore, the electron density can be determined as follows
\begin{equation} 
\label{siete_tito}
\frac{n_e}{n_0} = \frac{4}{\sqrt{\pi}} \left[ \int_{0}^{u_t} u^2 du + \int_{u_t}^{\infty}\exp -\left( u^2 - \phi\right)u^2 du \right],
\end{equation}
where we have defined $u = v/v_{\rm th}$. Using Eqs. (\ref{seis_tito}) and (\ref{siete_tito}), we can obtain a general expression for the electrostatic potential that casts the effects of the electron trapping
\begin{equation} 
\label{ocho_tito}
\nabla^2 \phi = \frac{1}{\lambda_{D}^{2}} \left( \frac{4}{3\sqrt{\pi}}\phi^{3/2} - f(\phi) - \exp \left( -\frac{r^2}{2\sigma_{0}^{2}} \right)\right),
\end{equation}
with $f(\phi)$ being given by
\begin{equation} 
\label{nueve_tito}
f(\phi) = \exp(\phi)\left( 1 - \frac{4}{\sqrt{\pi}}\int_{0}^{\sqrt{\phi}}\exp (-u^2) u^2du \right).
\end{equation}
Analytical solutions to Eq. (\ref{ocho_tito}) are not available in general. Fortunately, approximated expressions can be provided in some limiting cases. For weak trapping potentials, $\phi \ll 1$, we can use the expansion
\begin{equation}
f(\phi) \simeq 1 - \frac{4}{3\sqrt{\pi}}\phi^{3/2} + \phi - \frac{8}{15\sqrt{\pi}}\phi^{5/2},
\end{equation}
and the potential can be written as
\begin{equation} \label{diez_tito}
\nabla^2 \phi = \frac{1}{\lambda_{D}^{2}} \left[ 1 + \phi - \frac{8}{15\sqrt{\pi}}\phi^{5/2} - \exp \left( -\frac{r^2}{2\sigma_{0}^{2}} \right)\right].
\end{equation}
The other limiting case, which is more interesting in view of the experimental conditions, is the strong trapping regime, $\phi \gg 1$. In this case, $f(\phi) \simeq 0$ and the potential yields
\begin{equation} 
\label{once_tito}
\nabla^2 \phi = \frac{1}{\lambda_{D}^{2}} \left[ \frac{3}{4\sqrt{\pi}}\phi^{3/2} - \exp \left( -\frac{r^2}{2\sigma_{0}^{2}} \right)\right].
\end{equation}
In the strong confinement regime, we obtain an expression which is very similar to the Thomas-Fermi potential, used to describe the electronic distribution in heavy atomic species. The only difference is rooted in the fact that here the ions are not homogeneously distributed, and therefore we have included the ion inhomogeneity, which makes the model more suitable to
describe the physical situation of UNPs.
\begin{figure}[htp]
  \centering
  \caption{Snapshots of the spatial distribution of the plasma bubble. The blue (red) dots represent the electrons (ions). From top to bottom, $t=0$, $t = 0.99 /\omega_{pi}$ and $t= 2.1/\omega_{pi} $. We have set $N=10000$ and $\sigma_0=233 \mu$m, and the ions (electrons) are initiated at the temperature $T_e=95$ K ($T_i=1.3$ mK)}
  \label{fig_profiles}
\end{figure}

\section{Numerical Methods}

In order to confirm the previous model, we have performed molecular dynamics (MD) simulations of an expanding UNP, based on the Pretty Efficient Parallel Coulomb Solver (PEPC) code \cite{gibbon:2003:PEPC_parallelcoulomb}. The later combines the numerical techniques of the tree code method \cite{Nature.324.446} and fast multipole expansion \cite{JCompPhys.73.325}, allowing to reduce the computational cost from $N^2$ to $N \log N$ for $N$-particle systems. The dynamics of the system is determined by solving the following equations
\begin{equation}
 \label{uno_rolo}
\mathbf{r}_j(t + \delta t) = \mathbf{r}_j(t) + \mathbf{v}_j(t)\delta t + \frac{1}{2m_{i,e}}\mathbf{F}_j(t)\delta t^2,
\end{equation}
\begin{equation} 
\label{dos_rolo}
\mathbf{v}_j(t + \delta t) = \mathbf{v}_j(t) + \frac{\delta t}{2m_{i,e}}\left[ \mathbf{F}_j + \mathbf{F}_j(t + \delta t) \right],
\end{equation}
where
\begin{equation}
 \label{tres_rolo}
\mathbf{F}_j(t) = \sum _{i}^{j \neq i} \frac{1}{4\pi\varepsilon_0}\frac{q_jq_i\mathbf{r}_{ji}}{\left(r_{ji}^2 + \epsilon^2 \right)^{3/2}} 
\end{equation}
is the Coulomb force acting on the $j-$th particle at time $t$, and $\epsilon$ is a small exclusion radius to avoid numerical singularities. After checking the numerical sensitivity of the observables (e.g. the density profiles), we have set $\epsilon= a_{\rm WS}/100$, where $a_{\rm WS}=(3/4\pi n_0)^{1/3}$ stands for the Wigner-Seitz radius. Typical time steps $\delta t \ll \omega_{pe}$ have been set, small enough to resolve both the electron and ion dynamics. Moreover, we have chosen the ratio $m_i/m_e = 400$, in order to slow down the dynamics of the ions and still be able to study the effects of the ion distribution on the electrons. Lower values of $m_i/m_e$ generates a fast expansion of the ions, making it difficult to resolve the trapping effect; higher values imply the ions to be practically immobile at the electron time scales, demanding running the simulations for longer times and overcoming our present computational power. Finally, the temperature of the ions is obtained from the perpedicular component of the velocities, as
\begin{equation}
 \label{cuatro_rolo}
T_i = \frac{m_i}{2Nk_B} \sum_{i}^{N} \left[\frac{\left( \mathbf{v}\times \mathbf{r}\right)}{r} \right]^2,
\end{equation}
due to the fact that the kinetic energy is dominated by the radial expansion towards the edge of the plasma \cite{PhysRevE.81.046406}.

\section{Results and Discussion}

We have performed simulations with different number of particles in order to study different stages of electron trapping during the UNP expansion. We have distributed the electrons and the ions randomly, in agreement to the velocity distributions in Eq. (\ref{tres_tito}). The typical peak density $n_0 = 1 \times 10^8$ cm$^{-3}$ has been set for definiteness. The size of the cloud $\sigma_0$, depends on the number of particles, due to the fact the number of particles depends on the radial distribution function $n(r)4\pi r^2 dr$. The number of particles we used are $N_i = N_e = 10000$, $N_i = N_e = 20000$, $N_i = N_e = 30000$, which correspond to $\sigma_0 \simeq 233$ $\mu$m, $\sigma_0 \simeq 294$ $\mu$m and $\sigma_0 \simeq 336$ $\mu$m, respectively. These values are comparable to those obtained experimentally \cite{PhysRep.449.77}. In agreement with the typical experimental conditions, we have initialized the ions at the temperature $T_i\simeq 1.3$ mK, which leads to a coupling parameter 
\begin{equation}
\Gamma_i \equiv \frac{e^2}{4\pi\epsilon_0 a_{\rm WS}k_B T_i}\simeq 100.
\end{equation}
By doing so, we obtain a strongly correlated plasma but rule out Coulomb crystalization, occurring for $\Gamma_i\gtrsim 170$ \cite{PhysRevA.26.2255,RevModPhys.71.87}. Moreover, the initial electron temperature has been defined in such a way that Eq. (\ref{cuatro_tito}) is satisfied, which specifies the velocity of the electrons depending on their relative position inside the ion core. This criterion has been vastly employed in the study of stellar dynamics. We have followed Ref. \cite{AandA.37.183} to define the initial velocity distribution of the electrons, yielding $T_e \simeq 95$ K, $T_e \simeq 150$ K and $T_e \simeq 200$ K, for $N=10000$, $N=20000$ and $N=30000$, respectively. \par

In principal, three body recombination (TBR) could play a relevant role at low temperatures, being the main mechanism for electron heating \cite{PhysRevLett.86.3759, PhysRevLett.101.073202}. Classical TBR theory predicts the recombination rate per ion to value \cite{PhysRep.449.77}
\begin{equation}
K_{\rm TBR} \simeq 3.8 \times 10^{-9} \left(\frac{T_e}{{\rm K}}\right)^{-9/2} \left(\frac{n_e}{{\rm cm}^{-3}}\right)^2 \quad  {\rm s}^{-1}.
\end{equation}
In order to avoid TBR effects, an initial electron temperature above $40$ K is set in some experiments \cite{PhysPlasmas.22.033513}, a condition that we have observed above to be safely met at all stages of our simulations. As such, we have neglected TBR in the present work. \par 

In Fig. \ref{fig_profiles}, we show snapshots of the ionic and electronic radial distributions. After setting up the initial conditions, we let the system evolve according to Eq. \eqref{uno_rolo}. In our simulation, we have chosen a total simulation time $\tau = 180/\omega_{pe}\simeq 0.4$ $\mu$s (with $\omega_{pe}=\sqrt{n_e e^2/m_e\varepsilon_0}$ being the electron plasma frequency),  which is much larger than the Coulomb explosion time $\tau_{\rm exp} = \sqrt{m_i \sigma_{0}^{2}/k_B [T_e(0) + T_i(0)]}\simeq 122$ ns.\par
\begin{figure}[t!]
\centering
\subfigure{\includegraphics[width = 1\linewidth]{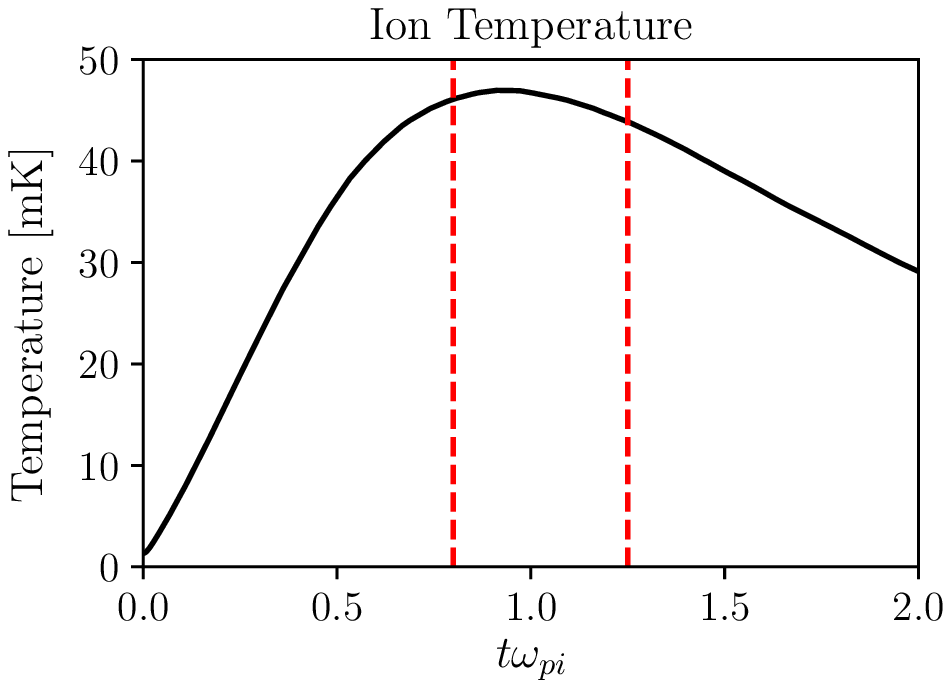}} \quad
\subfigure{\includegraphics[width = 1\linewidth]{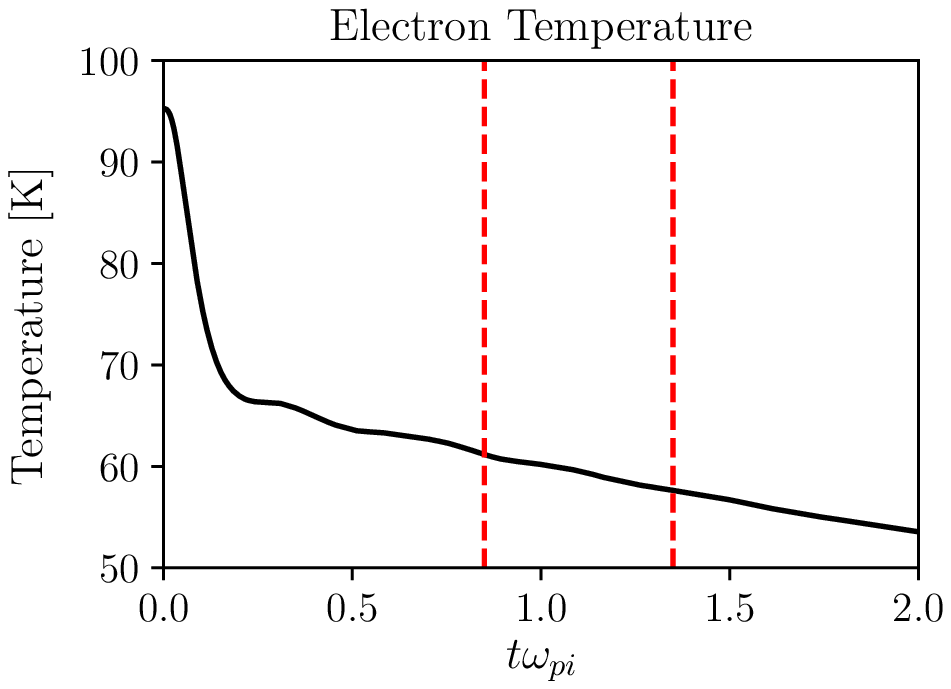}}
\caption{Electron and ion evolution of the average temperature with Gaussian initial distribution of the atoms for $N_i = N_e = 10000$ particles and $\sigma_{0} \approx 233$ $\mu$m. The initial temperatures have been set to $T_e \approx 95$ K and $T_i \approx 1.3$ mK. The time of the simulation is $t = 2 \omega_{p,i}^{-1}$ this is equivalent to $t = 50$ ns.}
\label{fig_temperatures}
\end{figure}

Fig. \ref{fig_temperatures} shows the evolution of the electron and ion temperatures. As we can see, the ion temperature increases, reaching its maximum at $t\sim \omega_{pi}^{-1}$. As we will discuss later on, this value is reached within the time window where the TF equilibrium is valid. After the TF period, the ions cool down again during the expansion. Conversely, the electronic temperature decreases monotonically, so the emergence of the TF equilibrium cannot be directly inferred from the temperature evolution. The same behavior is obtained (not shown) for different number of particles. \par 

\begin{figure}[b!]
\centering
\includegraphics[width =\linewidth]{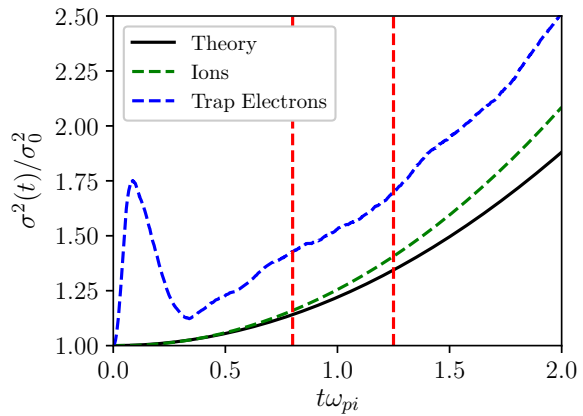}
\caption{Evolution of the mean square radius of the plasma cloud for $t = 2 \omega_{p,i}^{-1}$. The full line corresponds to the theoretical expression for the expansion of the ions and the dashed lines correspond to the values obtained with our simulation for $N_i = N_e = 20000$ particles and $\sigma_{0} = 294$ $\mu$m of initial radius for ions and electrons respectively.}
\label{fig_sigmas}
\end{figure}

We have also calculated the plasma expansion time by performing a typical Coulomb explosion analysis \cite{PhysRevLett.99.155001}. In the absence of correlations, the ion rms radius is given by $\sigma_{i}^2(t) = \sigma_0^2 (1 + t^2/\tau_{\rm exp}^2)$, where 
\begin{equation}
\tau_{\rm exp}= \left(\frac{m_i \sigma_{0}^{2}} {k_B (T_{e0} + T_{i0})}\right)^{1/2}
\end{equation}
is the expansion time. This is shown in Fig. \ref {fig_sigmas}, where we can see a good agreement between the theoretical model and the simulation for early stages. The ionic rms radius start to deviate from the theoretical prediction at the transition to the TF regime. This happens because the thermal electrons exert a drag force during the explosion. After the TF quasi-equilibrium is reached, electrons and ions expand at the same rate. \par 
\begin{figure}[t!]
  \centering
  \subfigure{\includegraphics[width = 1\linewidth]{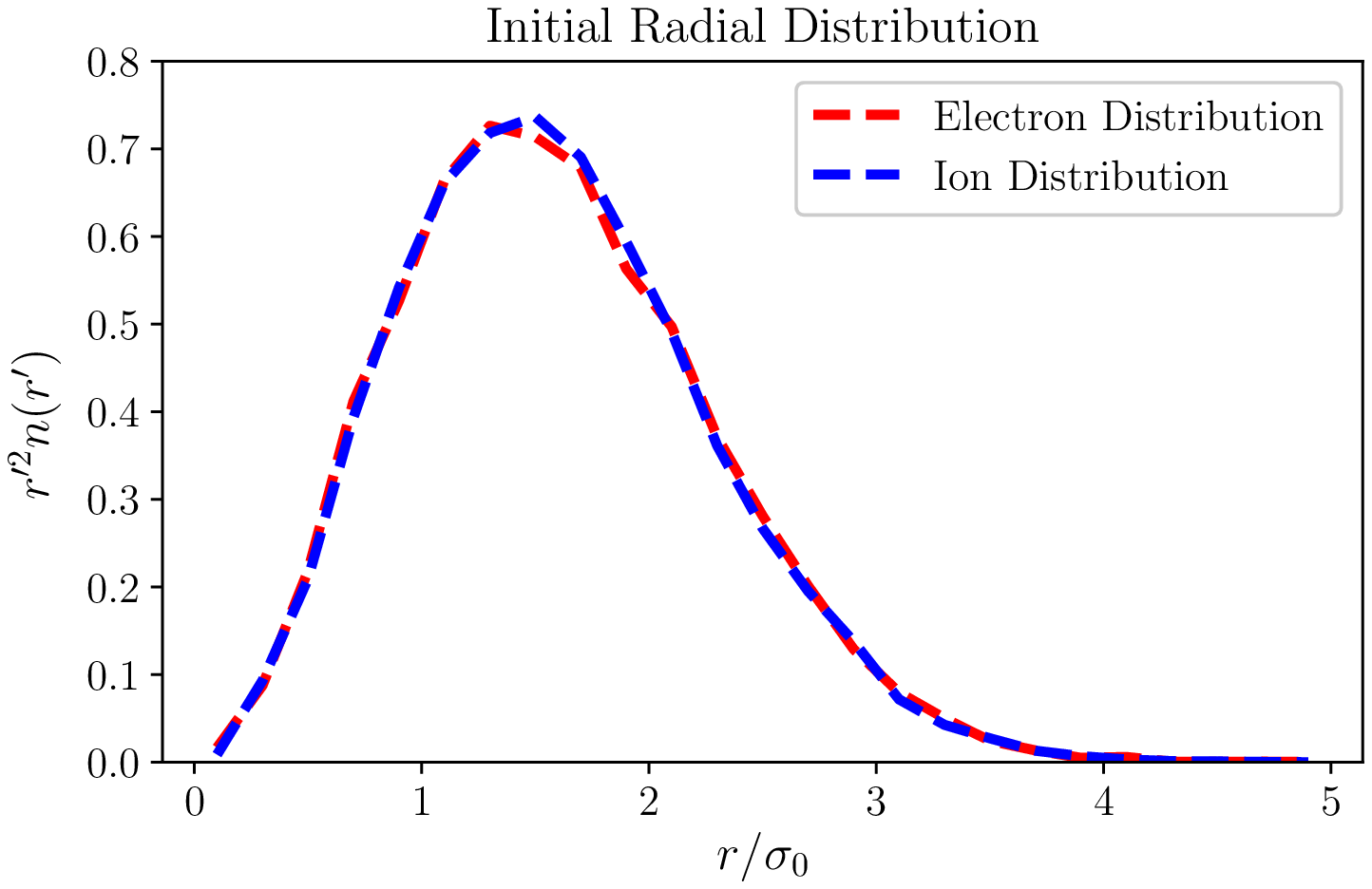}}\quad
  \subfigure{\includegraphics[width = 1\linewidth]{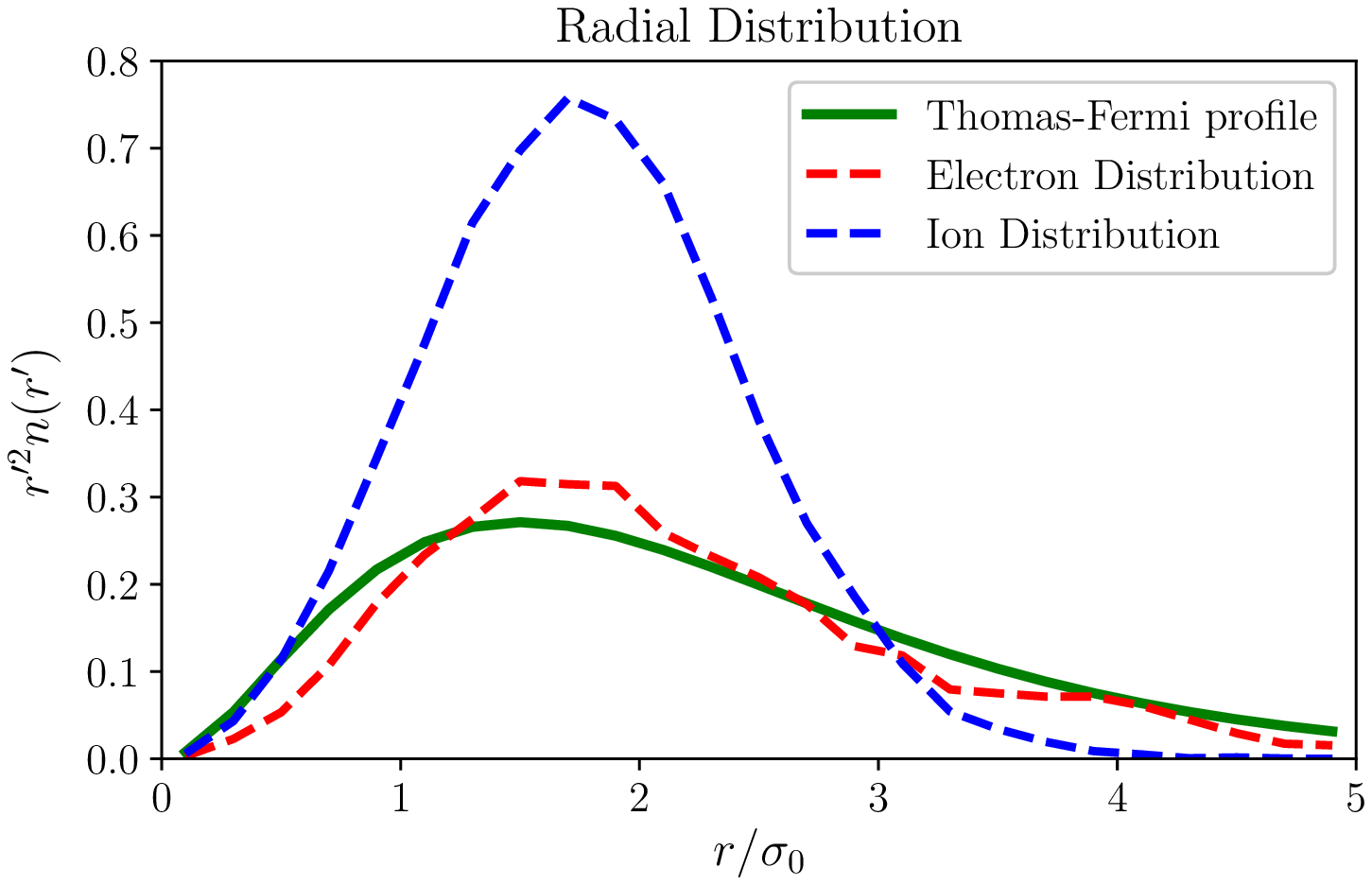}} \quad
  \subfigure{\includegraphics[width = 1\linewidth]{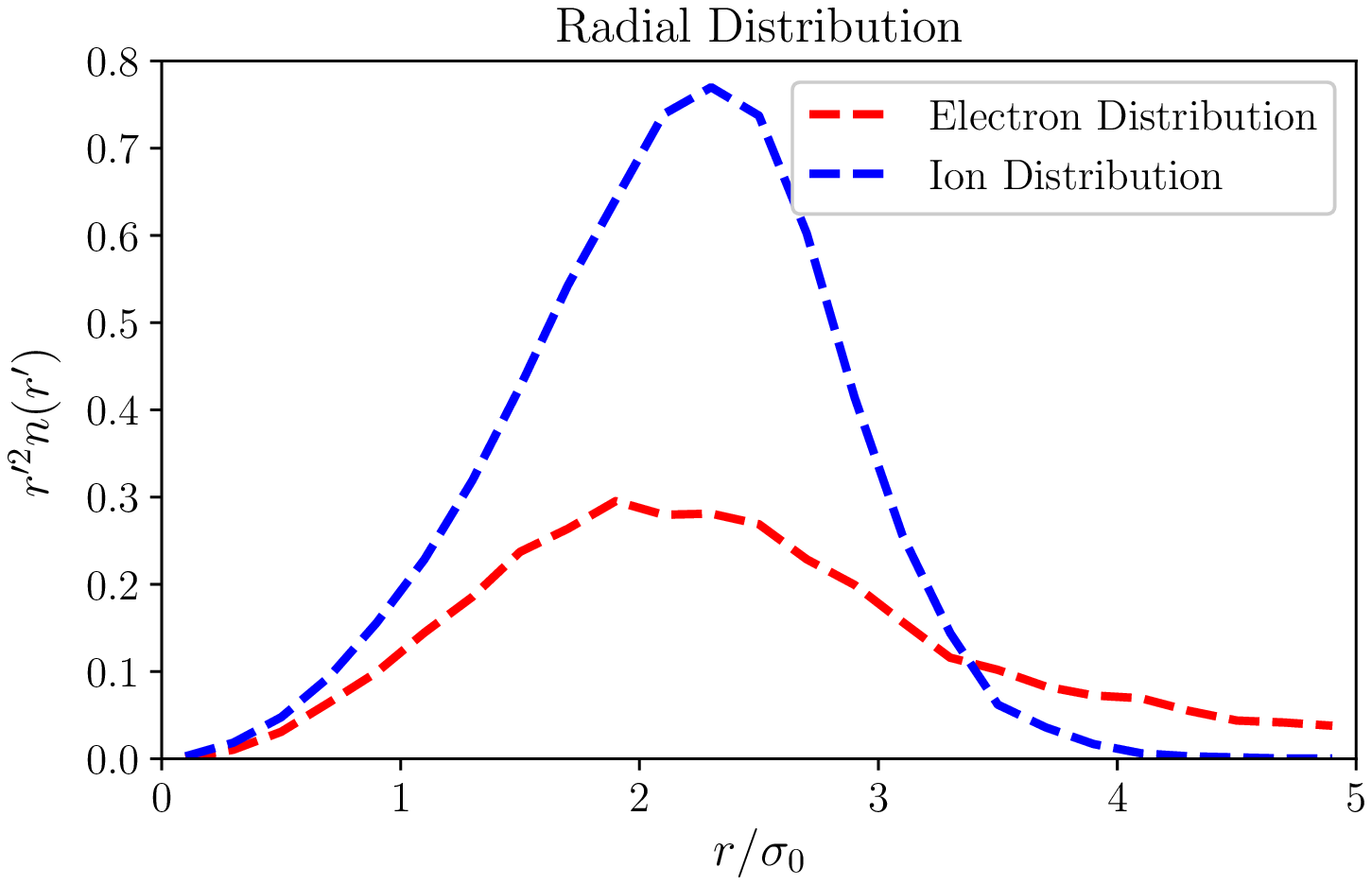}}
  \caption{Spatial distribution of ions and electrons in an ultracold neutral plasmas. The top panel corresponds to the initial gaussian distribution of particles with $N_i = N_e = 20000$ with $\sigma_0 = 294$ $\mu$m. The middle panel shows the corresponding distribution at $t = 1.01\omega_{p,i}^{-1}$, where the TF profile is good agreement with the simulation data. Finally, the bottom panel shows the distribution extracted at $t = 1.75\omega_{p,i}^{-1}$, where the TF equilibrium is no longer valid.}
  \label{fig_profiles2}
\end{figure}

\begin{figure}
  \centering
  \subfigure[]{\includegraphics[width = 0.48\linewidth]{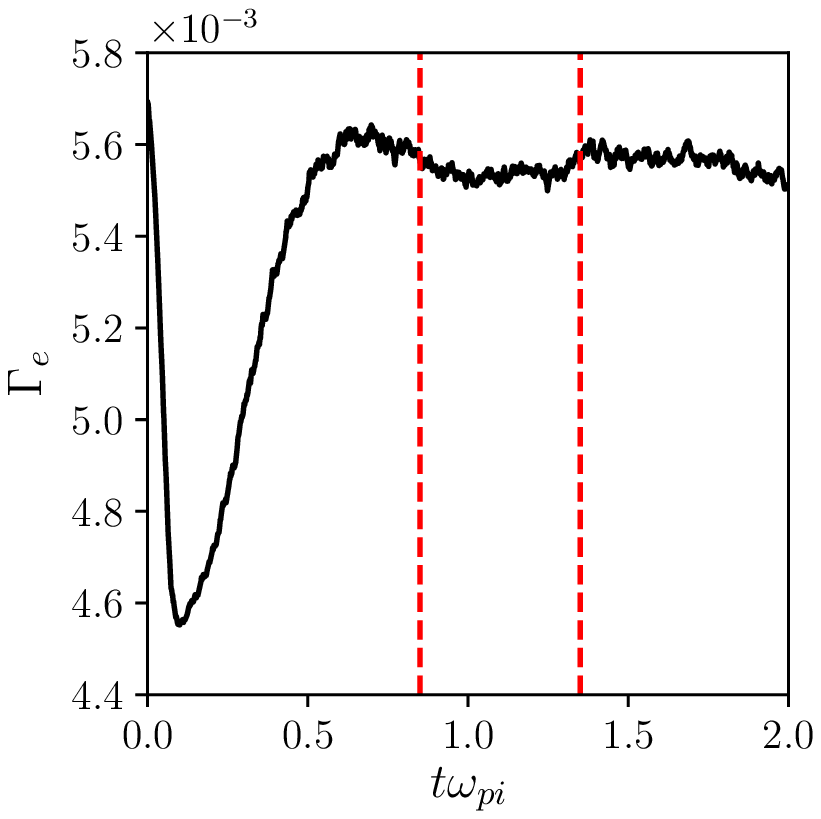}}
  \subfigure[]{\includegraphics[width = 0.48\linewidth]{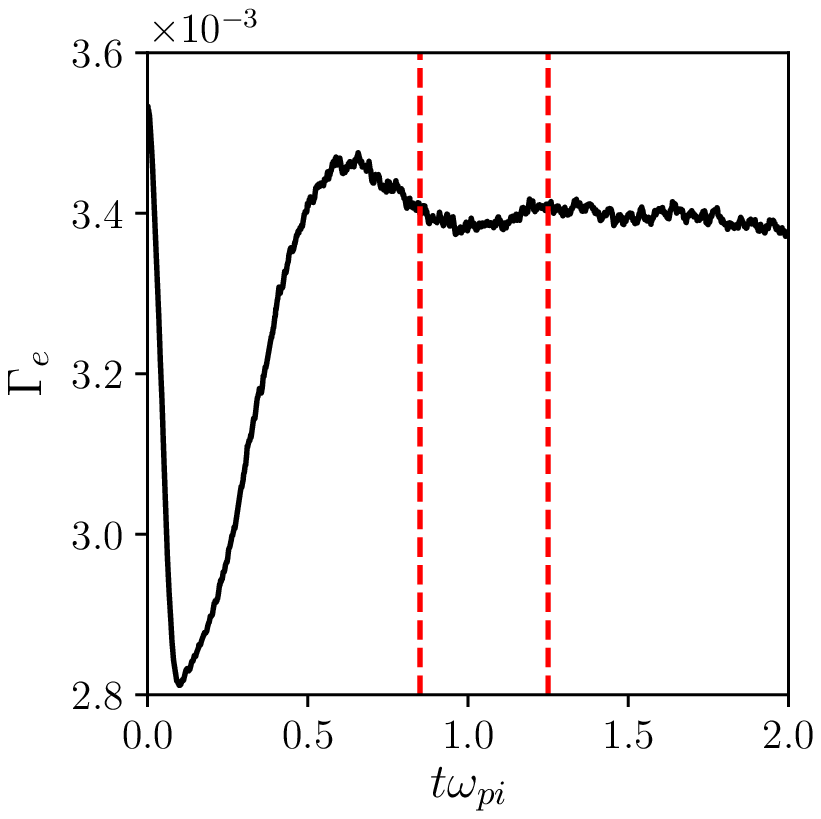}}
  \subfigure[]{\includegraphics[width = 0.48\linewidth]{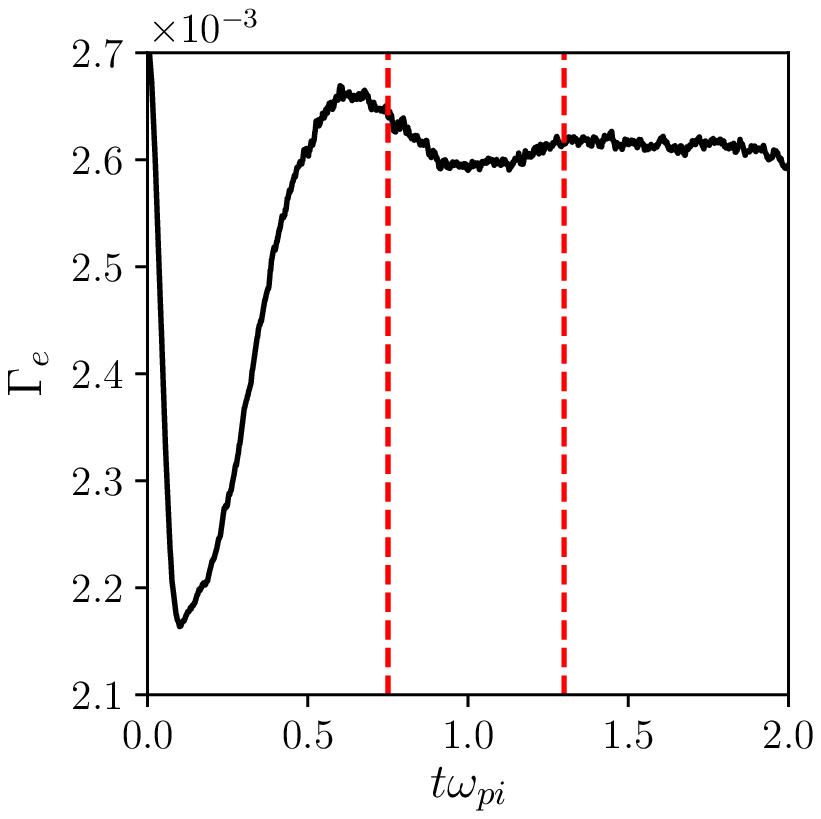}}
  \subfigure[]{\includegraphics[width = 0.48\linewidth]{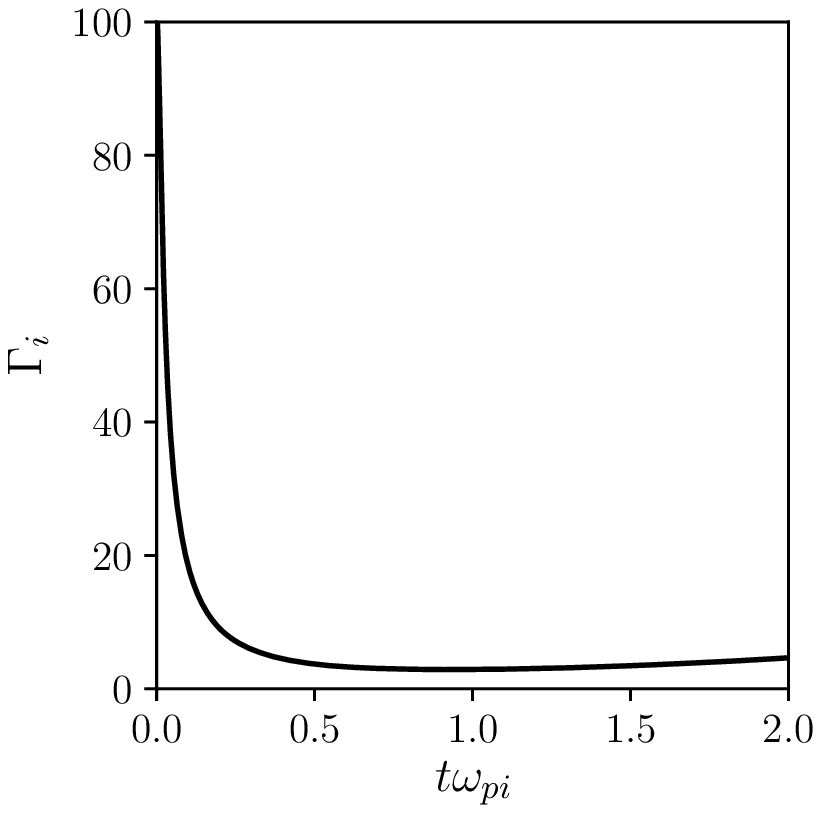}}
  \caption{Evolution of the electron and ion coupling parameters in the expanding UNPs, for a) $N_i = 10000$ and $\sigma_0 = 233$ $\mu$m, b) $N_i = 20000$ and $\sigma_0 = 294$ $\mu$m, and c) $N_i = 30000$ and $\sigma_0 = 336$ $\mu$m. We have fitted the distribution to the TF distribution in Eq. \eqref{once_tito}. Each of the snapshots are taken at $t = \omega_{p,i}^{-1}$. Panel d) depicts the evolution of the ion coupling parameter for the same parameters of panel a).}
  \label{fig_gammas}
\end{figure}

In order to relate the previous results to the establishment of a TF quasi-equilibrium, we compare the electron density profiles obtained from the MD simulations to those described by the theory. As depicted in Fig. \ref{fig_profiles2}, the electronic distribution is in quantitative agreement with the exponential profile of Eq. \eqref{siete_tito}, similarly to what happens with the electrons at the interior of heavy atomic species, for which the TF model holds. In the present case, the ions play the role of the nucleus and the electrons behave as the electronic cloud. According to our simulations, the TF quasi-equilibrium lasts for $\Delta t \simeq 0.6\omega_{pi}^{-1}= 12\omega_{pe}^{-1}$, as indicated by the vertical lines in Figs. \ref{fig_temperatures} and \ref{fig_sigmas}. After that time, the ion core is expanded enough for the electrons to become untrapped, resulting in the breakdown of the TF regime. 
Finally, we have computed the electronic coupling parameter $\Gamma_e$. A fast increase is observed at early stages, as a consequence of the Coulomb explosion of the electrons that leads to a fast electron depletion at the center. Once the TF regime is established, $\Gamma_e$ exhibits a minimum value. This is a signature of the electron trapping inside the ion cluster, where part of the kinetic energy is converted into the potential, thus increasing the electronic density locally. At later times, i.e. after the duration of the TF period, $\Gamma_e$ is kept almost constant. Although both the electronic temperature and density decrease during the expansion, we have $T_e \sim t^{-1}$, $n_e \sim t^{-3}$ during the expansion, yielding $\Gamma_e \sim {\rm const}$. These features are depicted in Fig. \ref{fig_gammas}.
\section{Conclusions}
In this work, we have studied the free expansion of a ultracold neutral plasma. Because of the temperature difference between ions and electrons, with the former being much lower than the latter, a quasi-equilibrium is formed due to the electron trapping that occurs inside the ion core. This transient regime is similar to the Thomas-Fermi model for heavy atomic species. The validity of our model is checked with the help of molecular dynamics calculations. We observe that the Thomas-Fermi lasts for a period of the order of the inverse of the ion plasma frequency, being followed by a regular Coulomb explosion afterwards.  

\section*{Acknowledgments}
The authors thanks FCT - Funda\c{c}\~{a}o da Ci\^{e}ncia e Tecnologia (Portugal) through the Ph.D. Grant PD/BD/105875/2014 (PD-F APPLAuSE) and through the grant number IF/00433/2015. Stimulating discussion with Jo\~ao Rodrigues are also acknowledged. 

\bibliography{References}

\begin{thebibliography}{26}%
\makeatletter
\providecommand \@ifxundefined [1]{%
 \@ifx{#1\undefined}
}%
\providecommand \@ifnum [1]{%
 \ifnum #1\expandafter \@firstoftwo
 \else \expandafter \@secondoftwo
 \fi
}%
\providecommand \@ifx [1]{%
 \ifx #1\expandafter \@firstoftwo
 \else \expandafter \@secondoftwo
 \fi
}%
\providecommand \natexlab [1]{#1}%
\providecommand \enquote  [1]{``#1''}%
\providecommand \bibnamefont  [1]{#1}%
\providecommand \bibfnamefont [1]{#1}%
\providecommand \citenamefont [1]{#1}%
\providecommand \href@noop [0]{\@secondoftwo}%
\providecommand \href [0]{\begingroup \@sanitize@url \@href}%
\providecommand \@href[1]{\@@startlink{#1}\@@href}%
\providecommand \@@href[1]{\endgroup#1\@@endlink}%
\providecommand \@sanitize@url [0]{\catcode `\\12\catcode `\$12\catcode
  `\&12\catcode `\#12\catcode `\^12\catcode `\_12\catcode `\%12\relax}%
\providecommand \@@startlink[1]{}%
\providecommand \@@endlink[0]{}%
\providecommand \url  [0]{\begingroup\@sanitize@url \@url }%
\providecommand \@url [1]{\endgroup\@href {#1}{\urlprefix }}%
\providecommand \urlprefix  [0]{URL }%
\providecommand \Eprint [0]{\href }%
\providecommand \doibase [0]{http://dx.doi.org/}%
\providecommand \selectlanguage [0]{\@gobble}%
\providecommand \bibinfo  [0]{\@secondoftwo}%
\providecommand \bibfield  [0]{\@secondoftwo}%
\providecommand \translation [1]{[#1]}%
\providecommand \BibitemOpen [0]{}%
\providecommand \bibitemStop [0]{}%
\providecommand \bibitemNoStop [0]{.\EOS\space}%
\providecommand \EOS [0]{\spacefactor3000\relax}%
\providecommand \BibitemShut  [1]{\csname bibitem#1\endcsname}%
\let\auto@bib@innerbib\@empty
\bibitem [{\citenamefont {Killian}(2007)}]{Science.316.705}%
  \BibitemOpen
  \bibfield  {author} {\bibinfo {author} {\bibfnamefont {T.~C.}\ \bibnamefont
  {Killian}},\ }\href {\doibase 10.1126/science.1130556} {\bibfield  {journal}
  {\bibinfo  {journal} {Science}\ }\textbf {\bibinfo {volume} {316}},\ \bibinfo
  {pages} {705} (\bibinfo {year} {2007})}\BibitemShut {NoStop}%
\bibitem [{\citenamefont {Killian}\ \emph {et~al.}(1999)\citenamefont
  {Killian}, \citenamefont {Kulin}, \citenamefont {Bergeson}, \citenamefont
  {Orozco}, \citenamefont {Orzel},\ and\ \citenamefont
  {Rolston}}]{PhysRevLett.83.4776}%
  \BibitemOpen
  \bibfield  {author} {\bibinfo {author} {\bibfnamefont {T.~C.}\ \bibnamefont
  {Killian}}, \bibinfo {author} {\bibfnamefont {S.}~\bibnamefont {Kulin}},
  \bibinfo {author} {\bibfnamefont {S.~D.}\ \bibnamefont {Bergeson}}, \bibinfo
  {author} {\bibfnamefont {L.~A.}\ \bibnamefont {Orozco}}, \bibinfo {author}
  {\bibfnamefont {C.}~\bibnamefont {Orzel}}, \ and\ \bibinfo {author}
  {\bibfnamefont {S.~L.}\ \bibnamefont {Rolston}},\ }\href {\doibase
  10.1103/PhysRevLett.83.4776} {\bibfield  {journal} {\bibinfo  {journal}
  {Phys. Rev. Lett.}\ }\textbf {\bibinfo {volume} {83}},\ \bibinfo {pages}
  {4776} (\bibinfo {year} {1999})}\BibitemShut {NoStop}%
\bibitem [{\citenamefont {Kulin}\ \emph {et~al.}(2000)\citenamefont {Kulin},
  \citenamefont {Killian}, \citenamefont {Bergeson},\ and\ \citenamefont
  {Rolston}}]{PhysRevLett.85.318}%
  \BibitemOpen
  \bibfield  {author} {\bibinfo {author} {\bibfnamefont {S.}~\bibnamefont
  {Kulin}}, \bibinfo {author} {\bibfnamefont {T.~C.}\ \bibnamefont {Killian}},
  \bibinfo {author} {\bibfnamefont {S.~D.}\ \bibnamefont {Bergeson}}, \ and\
  \bibinfo {author} {\bibfnamefont {S.~L.}\ \bibnamefont {Rolston}},\ }\href
  {\doibase 10.1103/PhysRevLett.85.318} {\bibfield  {journal} {\bibinfo
  {journal} {Phys. Rev. Lett.}\ }\textbf {\bibinfo {volume} {85}},\ \bibinfo
  {pages} {318} (\bibinfo {year} {2000})}\BibitemShut {NoStop}%
\bibitem [{\citenamefont {Killian}\ \emph {et~al.}(2001)\citenamefont
  {Killian}, \citenamefont {Lim}, \citenamefont {Kulin}, \citenamefont {Dumke},
  \citenamefont {Bergeson},\ and\ \citenamefont
  {Rolston}}]{PhysRevLett.86.3759}%
  \BibitemOpen
  \bibfield  {author} {\bibinfo {author} {\bibfnamefont {T.~C.}\ \bibnamefont
  {Killian}}, \bibinfo {author} {\bibfnamefont {M.~J.}\ \bibnamefont {Lim}},
  \bibinfo {author} {\bibfnamefont {S.}~\bibnamefont {Kulin}}, \bibinfo
  {author} {\bibfnamefont {R.}~\bibnamefont {Dumke}}, \bibinfo {author}
  {\bibfnamefont {S.~D.}\ \bibnamefont {Bergeson}}, \ and\ \bibinfo {author}
  {\bibfnamefont {S.~L.}\ \bibnamefont {Rolston}},\ }\href {\doibase
  10.1103/PhysRevLett.86.3759} {\bibfield  {journal} {\bibinfo  {journal}
  {Phys. Rev. Lett.}\ }\textbf {\bibinfo {volume} {86}},\ \bibinfo {pages}
  {3759} (\bibinfo {year} {2001})}\BibitemShut {NoStop}%
\bibitem [{\citenamefont {Killian}\ \emph {et~al.}(2007)\citenamefont
  {Killian}, \citenamefont {Pattard}, \citenamefont {Pohl},\ and\ \citenamefont
  {Rost}}]{PhysRep.449.77}%
  \BibitemOpen
  \bibfield  {author} {\bibinfo {author} {\bibfnamefont {T.}~\bibnamefont
  {Killian}}, \bibinfo {author} {\bibfnamefont {T.}~\bibnamefont {Pattard}},
  \bibinfo {author} {\bibfnamefont {T.}~\bibnamefont {Pohl}}, \ and\ \bibinfo
  {author} {\bibfnamefont {J.}~\bibnamefont {Rost}},\ }\href {\doibase
  https://doi.org/10.1016/j.physrep.2007.04.007} {\bibfield  {journal}
  {\bibinfo  {journal} {Physics Reports}\ }\textbf {\bibinfo {volume} {449}},\
  \bibinfo {pages} {77 } (\bibinfo {year} {2007})}\BibitemShut {NoStop}%
\bibitem [{\citenamefont {Mendon\ifmmode~\mbox{\c{c}}\else \c{c}\fi{}a}\ \emph
  {et~al.}(2008)\citenamefont {Mendon\ifmmode~\mbox{\c{c}}\else \c{c}\fi{}a},
  \citenamefont {Kaiser}, \citenamefont {Ter\ifmmode~\mbox{\c{c}}\else
  \c{c}\fi{}as},\ and\ \citenamefont {Loureiro}}]{PhysRevA.78.013408}%
  \BibitemOpen
  \bibfield  {author} {\bibinfo {author} {\bibfnamefont {J.~T.}\ \bibnamefont
  {Mendon\ifmmode~\mbox{\c{c}}\else \c{c}\fi{}a}}, \bibinfo {author}
  {\bibfnamefont {R.}~\bibnamefont {Kaiser}}, \bibinfo {author} {\bibfnamefont
  {H.}~\bibnamefont {Ter\ifmmode~\mbox{\c{c}}\else \c{c}\fi{}as}}, \ and\
  \bibinfo {author} {\bibfnamefont {J.}~\bibnamefont {Loureiro}},\ }\href
  {\doibase 10.1103/PhysRevA.78.013408} {\bibfield  {journal} {\bibinfo
  {journal} {Phys. Rev. A}\ }\textbf {\bibinfo {volume} {78}},\ \bibinfo
  {pages} {013408} (\bibinfo {year} {2008})}\BibitemShut {NoStop}%
\bibitem [{\citenamefont {Terças}\ \emph {et~al.}(2010)\citenamefont
  {Terças}, \citenamefont {Mendonça},\ and\ \citenamefont
  {Kaiser}}]{EurPhysLett.89.53001}%
  \BibitemOpen
  \bibfield  {author} {\bibinfo {author} {\bibfnamefont {H.}~\bibnamefont
  {Terças}}, \bibinfo {author} {\bibfnamefont {J.~T.}\ \bibnamefont
  {Mendonça}}, \ and\ \bibinfo {author} {\bibfnamefont {R.}~\bibnamefont
  {Kaiser}},\ }\href {http://stacks.iop.org/0295-5075/89/i=5/a=53001}
  {\bibfield  {journal} {\bibinfo  {journal} {EPL (Europhysics Letters)}\
  }\textbf {\bibinfo {volume} {89}},\ \bibinfo {pages} {53001} (\bibinfo {year}
  {2010})}\BibitemShut {NoStop}%
\bibitem [{\citenamefont {Morrison}\ \emph {et~al.}(2008)\citenamefont
  {Morrison}, \citenamefont {Rennick}, \citenamefont {Keller},\ and\
  \citenamefont {Grant}}]{PhysRevLett.101.205005}%
  \BibitemOpen
  \bibfield  {author} {\bibinfo {author} {\bibfnamefont {J.~P.}\ \bibnamefont
  {Morrison}}, \bibinfo {author} {\bibfnamefont {C.~J.}\ \bibnamefont
  {Rennick}}, \bibinfo {author} {\bibfnamefont {J.~S.}\ \bibnamefont {Keller}},
  \ and\ \bibinfo {author} {\bibfnamefont {E.~R.}\ \bibnamefont {Grant}},\
  }\href {\doibase 10.1103/PhysRevLett.101.205005} {\bibfield  {journal}
  {\bibinfo  {journal} {Phys. Rev. Lett.}\ }\textbf {\bibinfo {volume} {101}},\
  \bibinfo {pages} {205005} (\bibinfo {year} {2008})}\BibitemShut {NoStop}%
\bibitem [{\citenamefont {Saquet}\ \emph {et~al.}(2011)\citenamefont {Saquet},
  \citenamefont {Morrison}, \citenamefont {Schulz-Weiling}, \citenamefont
  {Sadeghi}, \citenamefont {Yiu}, \citenamefont {Rennick},\ and\ \citenamefont
  {Grant}}]{JPhysBAtMolOptPhys.44.184015}%
  \BibitemOpen
  \bibfield  {author} {\bibinfo {author} {\bibfnamefont {N.}~\bibnamefont
  {Saquet}}, \bibinfo {author} {\bibfnamefont {J.~P.}\ \bibnamefont
  {Morrison}}, \bibinfo {author} {\bibfnamefont {M.}~\bibnamefont
  {Schulz-Weiling}}, \bibinfo {author} {\bibfnamefont {H.}~\bibnamefont
  {Sadeghi}}, \bibinfo {author} {\bibfnamefont {J.}~\bibnamefont {Yiu}},
  \bibinfo {author} {\bibfnamefont {C.~J.}\ \bibnamefont {Rennick}}, \ and\
  \bibinfo {author} {\bibfnamefont {E.~R.}\ \bibnamefont {Grant}},\ }\href
  {http://stacks.iop.org/0953-4075/44/i=18/a=184015} {\bibfield  {journal}
  {\bibinfo  {journal} {J. Phys. B: At. Mol. Opt. Phys}\ }\textbf {\bibinfo
  {volume} {44}},\ \bibinfo {pages} {184015} (\bibinfo {year}
  {2011})}\BibitemShut {NoStop}%
\bibitem [{\citenamefont {Robicheaux}\ \emph {et~al.}(2014)\citenamefont
  {Robicheaux}, \citenamefont {Bender},\ and\ \citenamefont
  {Phillips}}]{JPhysBAtMolOptPhys.47.245701}%
  \BibitemOpen
  \bibfield  {author} {\bibinfo {author} {\bibfnamefont {F.}~\bibnamefont
  {Robicheaux}}, \bibinfo {author} {\bibfnamefont {B.~J.}\ \bibnamefont
  {Bender}}, \ and\ \bibinfo {author} {\bibfnamefont {M.~A.}\ \bibnamefont
  {Phillips}},\ }\href {http://stacks.iop.org/0953-4075/47/i=24/a=245701}
  {\bibfield  {journal} {\bibinfo  {journal} {J. Phys. B: At. Mol. Opt. Phys}\
  }\textbf {\bibinfo {volume} {47}},\ \bibinfo {pages} {245701} (\bibinfo
  {year} {2014})}\BibitemShut {NoStop}%
\bibitem [{\citenamefont {Fletcher}\ \emph {et~al.}(2007)\citenamefont
  {Fletcher}, \citenamefont {Zhang},\ and\ \citenamefont
  {Rolston}}]{PhysRevLett.99.145001}%
  \BibitemOpen
  \bibfield  {author} {\bibinfo {author} {\bibfnamefont {R.~S.}\ \bibnamefont
  {Fletcher}}, \bibinfo {author} {\bibfnamefont {X.~L.}\ \bibnamefont {Zhang}},
  \ and\ \bibinfo {author} {\bibfnamefont {S.~L.}\ \bibnamefont {Rolston}},\
  }\href {\doibase 10.1103/PhysRevLett.99.145001} {\bibfield  {journal}
  {\bibinfo  {journal} {Phys. Rev. Lett.}\ }\textbf {\bibinfo {volume} {99}},\
  \bibinfo {pages} {145001} (\bibinfo {year} {2007})}\BibitemShut {NoStop}%
\bibitem [{\citenamefont {Simien}\ \emph {et~al.}(2004)\citenamefont {Simien},
  \citenamefont {Chen}, \citenamefont {Gupta}, \citenamefont {Laha},
  \citenamefont {Martinez}, \citenamefont {Mickelson}, \citenamefont {Nagel},\
  and\ \citenamefont {Killian}}]{PhysRevLett.92.143001}%
  \BibitemOpen
  \bibfield  {author} {\bibinfo {author} {\bibfnamefont {C.~E.}\ \bibnamefont
  {Simien}}, \bibinfo {author} {\bibfnamefont {Y.~C.}\ \bibnamefont {Chen}},
  \bibinfo {author} {\bibfnamefont {P.}~\bibnamefont {Gupta}}, \bibinfo
  {author} {\bibfnamefont {S.}~\bibnamefont {Laha}}, \bibinfo {author}
  {\bibfnamefont {Y.~N.}\ \bibnamefont {Martinez}}, \bibinfo {author}
  {\bibfnamefont {P.~G.}\ \bibnamefont {Mickelson}}, \bibinfo {author}
  {\bibfnamefont {S.~B.}\ \bibnamefont {Nagel}}, \ and\ \bibinfo {author}
  {\bibfnamefont {T.~C.}\ \bibnamefont {Killian}},\ }\href {\doibase
  10.1103/PhysRevLett.92.143001} {\bibfield  {journal} {\bibinfo  {journal}
  {Phys. Rev. Lett.}\ }\textbf {\bibinfo {volume} {92}},\ \bibinfo {pages}
  {143001} (\bibinfo {year} {2004})}\BibitemShut {NoStop}%
\bibitem [{\citenamefont {McQuillen}\ \emph
  {et~al.}(2015{\natexlab{a}})\citenamefont {McQuillen}, \citenamefont
  {Castro}, \citenamefont {Bradshaw},\ and\ \citenamefont
  {Killian}}]{PhysPlasmas.22.043514}%
  \BibitemOpen
  \bibfield  {author} {\bibinfo {author} {\bibfnamefont {P.}~\bibnamefont
  {McQuillen}}, \bibinfo {author} {\bibfnamefont {J.}~\bibnamefont {Castro}},
  \bibinfo {author} {\bibfnamefont {S.~J.}\ \bibnamefont {Bradshaw}}, \ and\
  \bibinfo {author} {\bibfnamefont {T.~C.}\ \bibnamefont {Killian}},\ }\href
  {\doibase 10.1063/1.4918705} {\bibfield  {journal} {\bibinfo  {journal}
  {Physics of Plasmas}\ }\textbf {\bibinfo {volume} {22}},\ \bibinfo {pages}
  {043514} (\bibinfo {year} {2015}{\natexlab{a}})}\BibitemShut {NoStop}%
\bibitem [{\citenamefont {Chen}\ \emph {et~al.}(2016)\citenamefont {Chen},
  \citenamefont {Lu}, \citenamefont {Guo},\ and\ \citenamefont
  {Han}}]{PhysPlasmas.23.092102}%
  \BibitemOpen
  \bibfield  {author} {\bibinfo {author} {\bibfnamefont {T.}~\bibnamefont
  {Chen}}, \bibinfo {author} {\bibfnamefont {R.}~\bibnamefont {Lu}}, \bibinfo
  {author} {\bibfnamefont {L.}~\bibnamefont {Guo}}, \ and\ \bibinfo {author}
  {\bibfnamefont {S.}~\bibnamefont {Han}},\ }\href {\doibase 10.1063/1.4961957}
  {\bibfield  {journal} {\bibinfo  {journal} {Physics of Plasmas}\ }\textbf
  {\bibinfo {volume} {23}},\ \bibinfo {pages} {092102} (\bibinfo {year}
  {2016})}\BibitemShut {NoStop}%
\bibitem [{\citenamefont {Cummings}\ \emph {et~al.}(2005)\citenamefont
  {Cummings}, \citenamefont {Daily}, \citenamefont {Durfee},\ and\
  \citenamefont {Bergeson}}]{PhysRevLett.95.235001}%
  \BibitemOpen
  \bibfield  {author} {\bibinfo {author} {\bibfnamefont {E.~A.}\ \bibnamefont
  {Cummings}}, \bibinfo {author} {\bibfnamefont {J.~E.}\ \bibnamefont {Daily}},
  \bibinfo {author} {\bibfnamefont {D.~S.}\ \bibnamefont {Durfee}}, \ and\
  \bibinfo {author} {\bibfnamefont {S.~D.}\ \bibnamefont {Bergeson}},\ }\href
  {\doibase 10.1103/PhysRevLett.95.235001} {\bibfield  {journal} {\bibinfo
  {journal} {Phys. Rev. Lett.}\ }\textbf {\bibinfo {volume} {95}},\ \bibinfo
  {pages} {235001} (\bibinfo {year} {2005})}\BibitemShut {NoStop}%
\bibitem [{\citenamefont {Landau}\ and\ \citenamefont
  {Lifshitz}(1977)}]{LandauLifshitz}%
  \BibitemOpen
  \bibfield  {author} {\bibinfo {author} {\bibfnamefont {L.}~\bibnamefont
  {Landau}}\ and\ \bibinfo {author} {\bibfnamefont {E.}~\bibnamefont
  {Lifshitz}},\ }\href@noop {} {\emph {\bibinfo {title} {Quantum Mechanics}}},\
  \bibinfo {edition} {third edition}\ ed.\ (\bibinfo  {publisher} {Pergamon},\
  \bibinfo {year} {1977})\BibitemShut {NoStop}%
\bibitem [{\citenamefont {Gibbon}(2003)}]{gibbon:2003:PEPC_parallelcoulomb}%
  \BibitemOpen
  \bibfield  {author} {\bibinfo {author} {\bibfnamefont {P.}~\bibnamefont
  {Gibbon}},\ }\href@noop {} {\emph {\bibinfo {title} {PEPC : Pretty Efficient
  Parallel Coulomb-solver}}},\ \bibinfo {type} {Sonstiger Interner Bericht}\
  \bibinfo {number} {ZAM-IB-2003-05}\ (\bibinfo  {institution} {ZAM},\ \bibinfo
  {address} {J{\"{u}}lich, Forschungszentrum},\ \bibinfo {year}
  {2003})\BibitemShut {NoStop}%
\bibitem [{\citenamefont {Josh}\ and\ \citenamefont
  {Piet}(1986)}]{Nature.324.446}%
  \BibitemOpen
  \bibfield  {author} {\bibinfo {author} {\bibfnamefont {B.}~\bibnamefont
  {Josh}}\ and\ \bibinfo {author} {\bibfnamefont {H.}~\bibnamefont {Piet}},\
  }\href {\doibase 10.1038/324446a0} {\bibfield  {journal} {\bibinfo  {journal}
  {Nature}\ }\textbf {\bibinfo {volume} {324}},\ \bibinfo {pages} {446}
  (\bibinfo {year} {1986})}\BibitemShut {NoStop}%
\bibitem [{\citenamefont {Greengard}\ and\ \citenamefont
  {Rokhlin}(1987)}]{JCompPhys.73.325}%
  \BibitemOpen
  \bibfield  {author} {\bibinfo {author} {\bibfnamefont {L.}~\bibnamefont
  {Greengard}}\ and\ \bibinfo {author} {\bibfnamefont {V.}~\bibnamefont
  {Rokhlin}},\ }\href {\doibase http://dx.doi.org/10.1016/0021-9991(87)90140-9}
  {\bibfield  {journal} {\bibinfo  {journal} {Journal of Computational
  Physics}\ }\textbf {\bibinfo {volume} {73}},\ \bibinfo {pages} {325 }
  (\bibinfo {year} {1987})}\BibitemShut {NoStop}%
\bibitem [{\citenamefont {Guo}\ \emph {et~al.}(2010)\citenamefont {Guo},
  \citenamefont {Lu},\ and\ \citenamefont {Han}}]{PhysRevE.81.046406}%
  \BibitemOpen
  \bibfield  {author} {\bibinfo {author} {\bibfnamefont {L.}~\bibnamefont
  {Guo}}, \bibinfo {author} {\bibfnamefont {R.~H.}\ \bibnamefont {Lu}}, \ and\
  \bibinfo {author} {\bibfnamefont {S.~S.}\ \bibnamefont {Han}},\ }\href
  {\doibase 10.1103/PhysRevE.81.046406} {\bibfield  {journal} {\bibinfo
  {journal} {Phys. Rev. E}\ }\textbf {\bibinfo {volume} {81}},\ \bibinfo
  {pages} {046406} (\bibinfo {year} {2010})}\BibitemShut {NoStop}%
\bibitem [{\citenamefont {Slattery}\ \emph {et~al.}(1982)\citenamefont
  {Slattery}, \citenamefont {Doolen},\ and\ \citenamefont
  {DeWitt}}]{PhysRevA.26.2255}%
  \BibitemOpen
  \bibfield  {author} {\bibinfo {author} {\bibfnamefont {W.~L.}\ \bibnamefont
  {Slattery}}, \bibinfo {author} {\bibfnamefont {G.~D.}\ \bibnamefont
  {Doolen}}, \ and\ \bibinfo {author} {\bibfnamefont {H.~E.}\ \bibnamefont
  {DeWitt}},\ }\href {\doibase 10.1103/PhysRevA.26.2255} {\bibfield  {journal}
  {\bibinfo  {journal} {Phys. Rev. A}\ }\textbf {\bibinfo {volume} {26}},\
  \bibinfo {pages} {2255} (\bibinfo {year} {1982})}\BibitemShut {NoStop}%
\bibitem [{\citenamefont {Dubin}\ and\ \citenamefont
  {O'Neil}(1999)}]{RevModPhys.71.87}%
  \BibitemOpen
  \bibfield  {author} {\bibinfo {author} {\bibfnamefont {D.~H.~E.}\
  \bibnamefont {Dubin}}\ and\ \bibinfo {author} {\bibfnamefont {T.~M.}\
  \bibnamefont {O'Neil}},\ }\href {\doibase 10.1103/RevModPhys.71.87}
  {\bibfield  {journal} {\bibinfo  {journal} {Rev. Mod. Phys.}\ }\textbf
  {\bibinfo {volume} {71}},\ \bibinfo {pages} {87} (\bibinfo {year}
  {1999})}\BibitemShut {NoStop}%
\bibitem [{\citenamefont {Aarseth}\ \emph {et~al.}(1974)\citenamefont
  {Aarseth}, \citenamefont {Henon},\ and\ \citenamefont
  {Wielen}}]{AandA.37.183}%
  \BibitemOpen
  \bibfield  {author} {\bibinfo {author} {\bibfnamefont {S.~J.}\ \bibnamefont
  {Aarseth}}, \bibinfo {author} {\bibfnamefont {M.}~\bibnamefont {Henon}}, \
  and\ \bibinfo {author} {\bibfnamefont {R.}~\bibnamefont {Wielen}},\
  }\href@noop {} {\bibfield  {journal} {\bibinfo  {journal} {A\&A}\ }\textbf
  {\bibinfo {volume} {37}},\ \bibinfo {pages} {183} (\bibinfo {year}
  {1974})}\BibitemShut {NoStop}%
\bibitem [{\citenamefont {Bergeson}\ and\ \citenamefont
  {Robicheaux}(2008)}]{PhysRevLett.101.073202}%
  \BibitemOpen
  \bibfield  {author} {\bibinfo {author} {\bibfnamefont {S.~D.}\ \bibnamefont
  {Bergeson}}\ and\ \bibinfo {author} {\bibfnamefont {F.}~\bibnamefont
  {Robicheaux}},\ }\href {\doibase 10.1103/PhysRevLett.101.073202} {\bibfield
  {journal} {\bibinfo  {journal} {Phys. Rev. Lett.}\ }\textbf {\bibinfo
  {volume} {101}},\ \bibinfo {pages} {073202} (\bibinfo {year}
  {2008})}\BibitemShut {NoStop}%
\bibitem [{\citenamefont {McQuillen}\ \emph
  {et~al.}(2015{\natexlab{b}})\citenamefont {McQuillen}, \citenamefont
  {Strickler}, \citenamefont {Langin},\ and\ \citenamefont
  {Killian}}]{PhysPlasmas.22.033513}%
  \BibitemOpen
  \bibfield  {author} {\bibinfo {author} {\bibfnamefont {P.}~\bibnamefont
  {McQuillen}}, \bibinfo {author} {\bibfnamefont {T.}~\bibnamefont
  {Strickler}}, \bibinfo {author} {\bibfnamefont {T.}~\bibnamefont {Langin}}, \
  and\ \bibinfo {author} {\bibfnamefont {T.~C.}\ \bibnamefont {Killian}},\
  }\href {\doibase 10.1063/1.4915135} {\bibfield  {journal} {\bibinfo
  {journal} {Physics of Plasmas}\ }\textbf {\bibinfo {volume} {22}},\ \bibinfo
  {pages} {033513} (\bibinfo {year} {2015}{\natexlab{b}})}\BibitemShut
  {NoStop}%
\bibitem [{\citenamefont {Laha}\ \emph {et~al.}(2007)\citenamefont {Laha},
  \citenamefont {Gupta}, \citenamefont {Simien}, \citenamefont {Gao},
  \citenamefont {Castro}, \citenamefont {Pohl},\ and\ \citenamefont
  {Killian}}]{PhysRevLett.99.155001}%
  \BibitemOpen
  \bibfield  {author} {\bibinfo {author} {\bibfnamefont {S.}~\bibnamefont
  {Laha}}, \bibinfo {author} {\bibfnamefont {P.}~\bibnamefont {Gupta}},
  \bibinfo {author} {\bibfnamefont {C.~E.}\ \bibnamefont {Simien}}, \bibinfo
  {author} {\bibfnamefont {H.}~\bibnamefont {Gao}}, \bibinfo {author}
  {\bibfnamefont {J.}~\bibnamefont {Castro}}, \bibinfo {author} {\bibfnamefont
  {T.}~\bibnamefont {Pohl}}, \ and\ \bibinfo {author} {\bibfnamefont {T.~C.}\
  \bibnamefont {Killian}},\ }\href {\doibase 10.1103/PhysRevLett.99.155001}
  {\bibfield  {journal} {\bibinfo  {journal} {Phys. Rev. Lett.}\ }\textbf
  {\bibinfo {volume} {99}},\ \bibinfo {pages} {155001} (\bibinfo {year}
  {2007})}\BibitemShut {NoStop}%
\end{thebibliography}%

\end{document}